\documentclass[twocolumn,english]{revtex4-1}
\usepackage[T1]{fontenc}
\usepackage[latin9]{inputenc}
\usepackage{color}
\usepackage{textcomp}
\usepackage{amsmath}
\usepackage{graphicx}
\usepackage{esint}
\usepackage{babel}
\begin{document}
\title{Influence of toroidal flow on stationary density of collisionless
plasmas}
\author{Elias Laribi}
\affiliation{Aix Marseille Univ, Universit\'e de Toulon, CNRS, CPT, Marseille, France}
\affiliation{(current address) CEA, IRFM, F-13108 St. Paul-lez-Durance cedex, France}
\author{Shun Ogawa}
\affiliation{Laboratory for Neural Computation and Adaptation, RIKEN Center for
Brain Science, 2-1 Hirosawa Wako Saitama 351-0198, Japan}
\author{Guilhem Dif-Pradalier}
\affiliation{CEA, IRFM, F-13108 St. Paul-lez-Durance cedex, France}
\author{Alexei Vasiliev}
\affiliation{Space Research Institute, Profsoyuznaya 84/32, Moscow 117997, Russia}
\author{Xavier Garbet}
\affiliation{CEA, IRFM, F-13108 St. Paul-lez-Durance cedex, France}
\author{Xavier Leoncini}
\affiliation{Aix Marseille Univ, Universit\'e de Toulon, CNRS, CPT, Marseille, France}
\email{Xavier.Leoncini@cpt.univ-mrs.fr}

\begin{abstract}
Starting from the given passive particle equilibrium particle cylindrical profiles, we built self-consistent stationary conditions of the Maxwell-Vlasov equation at thermodynamic equilibrium with non-flat density profiles. The solutions to the obtained equations are then discussed. It appears that the presence of an azimuthal (poloidal) flow in the plasma can insure radial confinement, while the presence of a longitudinal (toroidal) flow can enhance greatly the confinement. Moreover in the global physically reasonable situation, we find that no unstable point can emerge in the effective integrable Hamiltonian of the individual particles, hinting at some stability of the confinement when considering a toroidal geometry in the large aspect ratio limit.\end{abstract}
\maketitle

\section{Introduction}

Insuring confinement of a hot plasma using a magnetic field, is one of the key issues to sustain in order to achieve magnetically confined fusion reactors. In these regards the emergence of a transport barrier that gives rise to the so-called H-mode has been a key ingredient in the design of most recent machines \citet{Wolf2003,Connor2004}. The study of these barriers has generated much work in the literature\citet{Balescu98,Firpo98,Ogawa2016_2}, but even though observed experimentally, up to our knowledge, there are currently no clear theoretical explanation of the emergence of such a barrier, neither has it really been observed as emerging from self-consistent numerical simulations (without some specially tailored external forcing). Following our recent study on the equilibrium solutions of passive particles described in \citet{Ogawa2019}, we built full self-consistent equilibrium solutions of the classical Maxwell-Vlasov equations. 
We then discuss what kind of motion can be expected in this self-consistent setting regarding the integrable motion of passive charged particles. We find out that in this setting no hyperbolic point emerges when looking at the effective potential driving the motion, meaning that in these type of cylindrical configurations, we do not expect a breaking of the magnetic moment when going back to a toroidal geometry due to separatrix crossings such as the one displayed in \citet{Cambon2014} or the presence of Hamiltonian chaos due to this mechanism \citet{Neishtadt86,Tennyson86,Cambon2014,Ogawa2016,Leoncini2018}. However since we have a full family of solutions, we can as well study the influence of the presence of some ``toroidal'' flow in the plasma, on the plasma density profile, and we show it can enhance plasma confinement. Our approach stems from first principles, so it is considered in a very idealized situation, but we believe it could be useful and shed possible light on the origin of improved plasma confinement, such as the rise of transport barriers and how to trigger them. This paper is organized as follows, in the first part we present the idealized configuration and the passive particle thermal equilibrium from \citet{Ogawa2019}, then building from this we discuss the self-consistent approach with one species, and finally move on to a real full self-consistent solution with two species.

\section{Plasma setting and passive equilibrium}

Let us recall the conditions that were discussed in \citet{Ogawa2019}. 
We consider an infinite aspect ratio limit of the tokamak torus such that we end up with a cylindrical configuration In this configuration, we choose an ideal magnetic field of the form
\[
\overrightarrow{B(r)}=B_{0}\hat{e}_{z}+B_{0}g(r)\hat{e}_{\theta}\:,
\]
whose vector potential field in a Coulomb gauge writes. 
\[
\overrightarrow{A(r)}=\frac{B_{0}r}{2}\hat{e}_{\theta}-B_{0}F(r)\hat{e}_{z}\:,
\]
where $\hat{e}_{z}$ is the unit vector along the cylinder axis (corresponding to the toroidal direction in the torus), $\hat{e}_{\theta}$ the unit vector on the poloidal direction, 
$B_{0}$ is the intensity of the uniform magnetic field (assumed to be generated by external coils), and $F$ is the primitive of $g$, that can be linked to the plasma current $j_{z}(r)$ or the so called $q$-profile
(introducing some fictitious characteristic radius $R_{\rm per}$) by
\[
F(r)=\intop^{r}\frac{rdr}{R_{\rm per}q(r)}=\int^{r}\frac{j_{z}(r)r}{B_{0}}dr\:.
\]
We shall consider a positive charged particle evolving in such magnetic filed, without any electric field. 
The motion of the particle is then described by the following effective one degree of freedom Hamiltonian
\begin{equation}
H=\frac{1}{2m}\left[p_{r}^{2}+\left(\frac{p_{\theta}}{r}-\frac{qB_{0}r}{2}\right)^{2}+\left(p_{z}-qA_{z}(r)\right)^{2}\right]\:.\label{eq:Hamiltonien}
\end{equation}
Indeed due to the symmetries of the problem (translation along $z$ and rotation of $\theta$ around the cylinder axis), we have on top
of the kinetic energy of the system (the Hamiltonian $H$.) two extra conserved quantities $p_{\theta}$ and $p_{z}$, turning this 3-degree
of freedom problem into an integrable one with an effective one-degree of freedom Hamiltonian. Here we have an effective potential energy that
depends on the initial condition through the constant values of $p_{\theta}$ and $p_{z}$. We can note as well that the effective potential depends
on the poloidal magnetic field through the function $F(r)$, it is then possible to imagine settings that give rise to a hyperbolic point and an associated separatrix. 
The presence of these separatrices in the full phase space can then after some perturbation be the natural root of separatrix chaos and the breaking of adiabatic invariants
through separatrix crossings. We shall come back to this statement, as this feature was shown to break the magnetic moment in some regions of the full phase space \citet{Cambon2014}.

Let us now built a passive particle equilibrium distribution from the previous considerations. Using the Hamiltonian $(\ref{eq:Hamiltonien}$),
we can find a stationary one particle-density function obeying some Vlasov equation \citet{Jeans1915,Vlasov38,Vlasov68,Binney2008} we just need to find $f(p_{r},p_{\theta},p_{z},r,\theta,z,t)$ such that
\[
\frac{Df}{Dt}=0\:,
\]
where we used the particle derivative with the specificity that 
\[
\frac{\partial f}{\partial t}=0\:.
\]
Solutions of this problem satisfy necessarily the following equation
\[
\left\{ f,H\right\} =0\:,
\]
where $\left\{ \cdot,\cdot\right\} $ denotes the Poisson brackets.
Any function $f$ that is just a function of $H$ is a solution of the problem. 
In order to choose one among the infinite possibility, we applied the maximum entropy principle \citet{Jaynes57,Zubarev1996,Yamaguchi04}, 
imposing constraints due to the different invariants of the dynamics coming from the symmetries of the problem and the total number of particles in the system. 
These lead to the introduction of three Lagrangian multipliers $\beta$, $\gamma_{\theta}$, $\gamma_{z}$ respectively associated to the energy, the angular momentum and the translational invariance. 
We then obtain as a result the following density function,
\begin{equation}
f\propto e^{-\beta H-\gamma_{\theta}p_{\theta}-\gamma_{z}p_{z}}\:,\label{eq:Station_passive_soluition}
\end{equation}
$\gamma_{1}$ the last multiplier dealing with the normalization of $f$ (i.e conservation of number of particles $N$, $\int fd^{3}\mathbf{p}d^{3}\mathbf{q}=N$),
is absorbed in the proportional coefficient of (\ref{eq:Station_passive_soluition}).

\noindent From this stationary distribution we can get for instance the spatial density $n(\mathbf{q})$ 
as 
\begin{equation}
n(\mathbf{q})\equiv\int f_{0}d^{3}\mathbf{p}=\int f_{0}r^{-1}dp_{\theta}dp_{z}dp_{r}.
\end{equation}
We get equivalently the charge density by 
\begin{equation}
\rho(r)=\frac{Nq\exp\left(-ar^{2}-bA_{z}(r)\right)}{4\pi^{2}R_{\rm per}\int_{0}^{{\color{red}{\normalcolor \infty}}}r\exp\left(-ar^{2}-bF(r)\right)dr},\label{eq:density}
\end{equation}
where 
\[
a=\frac{\gamma_{\theta}}{2}\left(qB_{0}-\frac{m\gamma_{\theta}}{\beta}\right),\quad b=\gamma_{z}qB_{0}.
\]
different situations may arise in this profile and we refer to  \citet{Ogawa2019} for some of the results and conclusions. Let us  insist on the fact that the Lagrangian multipliers $\gamma_z$ is due to the  conservation of $p_z$, and is directly linked to its average. A non zero $\gamma_z$ corresponds therefore to the existence of some global movement of the plasma, and mass flow. Of course, as shall be discussed later, motion of ions induces as well a current, but it is important to recall that if there is no global flow ($\gamma_z=0$ and $\gamma_\theta=0$), then the equilibrium is trivial and there is just a flat profile, and non-flat profiles in an non flowing plasma correspond then to an out of equilibrium states.
But now let us move to the construction of a self-consistent solution of the problem.

\section{Self-consistent equation}

\subsection{One species in neutralizing background\label{subsec:One-species-in}}

Before moving to self-consistency, we will be more specific in what is meant by self-consistent. We consider that the electric and magnetic fields are created by the particles, using Gauss and Amp\'ere laws,
meaning we assume that those fields are static and created by the stationary distribution, we as well consider non-relativistic particles.
In order to built on what has been done before, we consider  first a system with one specied of charged particles and without electric field. 
In other words there is some static background that neutralizes the charge density. 
Moreover we will consider that the component of the magnetic field along the cylinder axis $B_{0}\hat{e}_{z}$ is generated by some external coils.

If one now looks back at the passive particle distribution one notice that it is possible as well to compute the density of current flowing along the axis by performing the integration
\noindent 
\[
j_{z}(r)=q\int v_{z}f_{0}d^{3}p, 
\]
where
\[
v_{z}=\frac{p_{z}-qA_{z}(r)}{m}, 
\]
and we obtain
\begin{equation}
	\begin{split}
	j_{z}(r) & =\frac{-Nq\gamma_{z}\exp\left(-ar^{2}-bA_{z}(r)\right)}{4\pi^{2}R_{\rm per}\beta\int_{0}^{{\color{red}{\normalcolor \infty}}}r\exp\left(-ar^{2}-bF(r)\right)dr}\:.\label{eq:Current}\\
	 & =-\frac{\gamma_{z}}{\beta}\rho(r)
	\end{split}	
\end{equation}
To move to  a partial self-consistent setting, we neglect the influence of the poloidal current on the axial component of the magnetic field which will remain $B_{0}\hat{e}_{z}$. This can be for instance justified when $B_0$ is large and plasma density low, we may consider for instance when $q B_0\gg m\gamma_\theta/\beta)$ such that $a$ is linear in $\gamma_\theta$, however having in mind an analogy with the tokamak, the poloidal magnetic field is generated by the current of the particles, we shall have
\begin{equation}
\Delta A_{z}=-\mu_{0}j_{z}(r)\:,\label{eq:Poisson_Eq}
\end{equation}
so we end up with an implicit equation in the vector potential which if it has solutions leads to self-consistent solutions of the Vlasov system, that explicitly writes
\[
\frac{1}{r}\frac{d}{dr}\left(r\frac{dA_{z}(r)}{dr}\right)\propto\gamma_{z}e^{-ar\text{\texttwosuperior}-bA_{z}(r)}\:.
\]
Removing all the dimensions and multiplicative terms by rescaling, we end up with an equation,
\begin{equation}
\frac{1}{\bar{r}}\frac{d}{d\bar{r}}\left(\bar{r}\frac{d\psi(\bar{r})}{d\bar{r}}\right)=-e^{-\bar{a}\bar{r}^{2}+\psi(\bar{r})}\:,\label{eq:implicit_sans_dim}
\end{equation}
where we introduced variable $\bar{r}$ and the function $\psi(\bar{r})$. 
For simplicity we now will get rid of the $\bar{}$ term in the following.
We can notice that, when making the change of variables in the end the sign of $\gamma_{z}$ becomes irrelevant as it introduces a dependence on $\gamma_{z}^{2}$. 
This is not really surprising as the end result should not depend on how the direction of the plasma current flows, for the sake of simplicity we will assume in the rest of the paper that $\gamma_{z}>0$.

\subsection{Full self-consistent solution}

To find a full self-consistent solution of the Vlasov system, we do not assume a neutralizing background, 
in order to still be coherent we then need to add a second type of particles whose charges have an opposite charge to the initial ones (for instance electrons if we had protons or alpha particles). 
Since we have a choice, we shall consider that the previous equilibrium was computed for positive charges ($q>0$). 
Then for the neutralizing negative particles, we will follow the same procedure and maximize each entropy of each type of particles independently (this hypothesis could be discussed, 
in someway we are assuming no collisions, even if we will need to impose electroneutrality). 
We will use the $\pm$ sign to refer for instance to the distribution of positive and negative particles, so the previous expressions, 
for instance $\rho(r)$ from equation (\ref{eq:density}) will be noted $\rho^{+}$(and multiplied by the charge of the particle to become the positive charge density), 
and the same for the current $j_{z}$ from (\ref{eq:Current}) will become $j_{z}^{+}$. 
So lets us compute $f^{-}$, $\rho^{-}(r)$, $j_{z}^{-}(r)$, the Hamiltonian of the passive negative particles writes
\[
H^{-}=\frac{\left(p_{r}^{-}\right)^{2}}{2m^{-}}+\frac{\left(\frac{p_{\theta}^{-}}{r}+\frac{qB_{0}r}{2}\right)^{2}}{2m^{-}}+\frac{\left(p_{z}^{-}+qA_{z}(r)\right)^{2}}{2m^{-}}\:,
\]
which lead to the passive distribution 
\[
f^{-}\propto e^{-\beta^{-}H^{-}-\gamma_{\theta}^{-}p_{\theta}^{-}-\gamma_{z}^{-}p_{z}^{-}}
\]
and density of negative charges
\[
\rho^{-}(r)=\frac{-qe^{-a^{-}r^{2}-b^{-}A_{z}(r)}}{4\pi^{2}R_{\rm per}\int re^{-a^{-}r^{2}-b^{-}A_{z}(r)}dr}
\]
 with $a^{-}=-\frac{\gamma_{\theta}^{-}}{2}\left(qB_{0}+\frac{m^{-}\gamma_{\theta}^{-}}{\beta^{-}}\right)$,
$b^{-}=-q\gamma_{z}^{-}$. We can as well in order to move to self
consistency, compute the current induced by these negative charges
\[
j_{z}^{-}(r)=\frac{q\gamma_{z}^{-}e^{-a^{-}r\text{\texttwosuperior}-b^{-}A_{z}(r)}}{4\pi\text{\texttwosuperior}R_{\rm per}\beta^{-}\int re^{-a^{-}r\text{\texttwosuperior}-b^{-}A_{z}(r)}dr}\:,
\]
And we end up with the self-consistent Poisson type equation that
writes:
\[
\frac{1}{r}\frac{d}{dr}\left(r\frac{dA_{z}(r)}{dr}\right)=-j_{z}^{+}(r)-j_{z}^{-}(r)
\]
and since we assumed no electric field, we as well have to impose
electro-neutrality:
\begin{equation}
\rho^{-}(r)+\rho^{+}(r)=0\:,\label{eq:electroneutralite}
\end{equation}
 which leads to 
\begin{equation}
-C_{-}e^{-a^{-}r^{2}-b^{-}A_{z}(r)}+C_{+}e^{-a^{+}r^{2}-b^{+}A_{z}(r)}=0\label{eq:electorneutralite2}
\end{equation}
 where 
\[
C_{\pm}=\frac{Nq}{4\pi\text{\texttwosuperior}R_{\rm per}\beta^{\pm}\int re^{-a^{\pm}r\text{\texttwosuperior}-b^{\pm}A_{z}(r)}dr}\:.
\]

\noindent Let us now discuss the possible solutions.
\begin{itemize}
\item The first case to consider is if $b^{-}\neq b^{+}$, the equation
(\ref{eq:electorneutralite2}) then implies that $A_{z}$ can eventually
depend on $r^{2}$ 
\[
A_{z}(r)=\frac{log\left(\frac{C_{-}}{C_{+}}\right)}{(b^{-}-b^{+})}+\left(\frac{a^{+}-a^{-}}{b^{-}-b^{+}}\right)r^{2}
\]
and that we end up with a uniform constant current $j_{z}$. This
is not really interesting physically..
\item The other case of possible non-trivial solution which satisfy self
consistency and electroneutrality implies $b^{-}=b^{+}=b$ and therefore
$a^{-}=a^{+}=a$ and $C_{+}=C_{-}=C$, this leads to a non-uniform
current 
\[
j_{z}(r)=-Cq\gamma_{z}^{+}e^{-ar^{2}-bA_{z}(r)}\left(\frac{1}{\beta^{+}}+\frac{1}{\beta^{-}}\right)
\]
 with possibly the two species at different temperatures. The self-consistent
equations then ends up writing as
\[
\frac{1}{r}\frac{d}{dr}\left(r\frac{dA_{z}(r)}{dr}\right)=Cq\gamma_{z}^{+}e^{-ar^{2}-bA_{z}(r)}\left(\frac{1}{\beta^{+}}+\frac{1}{\beta^{-}}\right)
\]
And if we get rid of the dimensions, we end up with a formally identical
equation as (\ref{eq:implicit_sans_dim}) ,
\[
\frac{1}{\bar{r}}\frac{d}{d\bar{r}}\left(\bar{r}\frac{d\psi(\bar{r})}{d\bar{r}}\right)=-e^{-\bar{a}\bar{r}^{2}+\psi(\bar{r})}\:.
\]
\end{itemize}
In this context adding a second species ends up being in solving the
same implicit equation, but we where then able to get rid of the neutralizing
background assumption and potentially be more physically relevant.

\section{Solutions of the self-consistent equation}

In order to solve the equation (\ref{eq:implicit_sans_dim}) we can
envision different situations, namely $a>0$, $a=0$ and eventually
$a<0$. To tackle the problem we can as well make some transformations,
if we note $\phi(r)=-ar^{2}+\psi(r)$, we end up with
\begin{equation}
\Delta\phi(r)+4a=-e^{\phi(r)}\:,
\end{equation}
but it is not much simpler. We now assume that $a>0$, so we will
have to deal with a $\pm$ sign if the initial $a\ne0$. We can then
rescale the length again and obtain
\begin{equation}
\Delta\phi(r)\pm1=-e^{\phi(r)-\log(4a)}\:,
\end{equation}
and then shift $\phi$ to arrive at
\begin{equation}
\Delta\phi(r)\pm1=-e^{\phi(r)}\:.\label{eq:Adimensionalized_self_conis}
\end{equation}
Since $r>0$, we may use the notation $t=\log r$, and we end up with 
\[
\ddot{\phi}=-e^{\phi+2t}\mp e^{2t}\:,
\]
where the $\dot{}$ refers to $d/dt$ and if we note $\varphi=\phi+2t$
(in some way we change our reference frame), we finally obtain
\begin{equation}
\ddot{\varphi}=-e^{\varphi}\mp e^{2t}\:,\label{eq:Newton1}
\end{equation}
where we can recognize some newton equation, with a force deriving
from a potential and one that is just time dependent (when $a=0$
the time dependent force is simply zero and we have an autonomous
system).

\subsection{Bennett-pinch solution when $a=0$}

A family of solutions of this equation can be found in the literature
under the name of Bennet pinch solution \citet{Bennett34,Bennett55},
it occurs when we consider a distribution with $a=0$. In fact, the
Bennett pinch corresponds to a particular case of the general solutions
of the Lane Emden like equation when for $a=0$ which means $\gamma_{\theta}=0$
or $\gamma_{\theta}=\frac{qB_{0}\beta}{m}$.
\begin{equation}
\Delta\psi(r)=-e^{\psi(r)},
\end{equation}
which corresponds to the use of any holomorphic function $w$ as 
\begin{equation}
e^{\psi(r)}=\frac{8|w'|^{2}}{(1+|w|^{2})^{2}},\quad{\rm where}\quad w'=\frac{{\rm d}w}{{\rm d}z}.
\end{equation}
The function $w$ should holds that both $\|w\|$ and $\|w'\|$ depend
only on $r=|z|$. For  more details a good starting point can be found in \citet{Cecherini05}. A trivial example is that $w=cr^{\alpha}$. In this
case, 
\begin{equation}
e^{\psi(r)}=\frac{8c^{2}\alpha^{2}r^{2\alpha-2}}{(1+c^{2}r^{2\alpha})^{2}}.
\end{equation}
and the Bennet pinch solution corresponds to the choice $\alpha=1$:
\begin{equation}
e^{\psi(r)}=\frac{8c^{2}}{\left(1+c^{2}r^{2}\right)^{2}}.\label{eq:Bennet-pinch}
\end{equation}

\begin{figure}
\begin{centering}
\includegraphics[width=8cm]{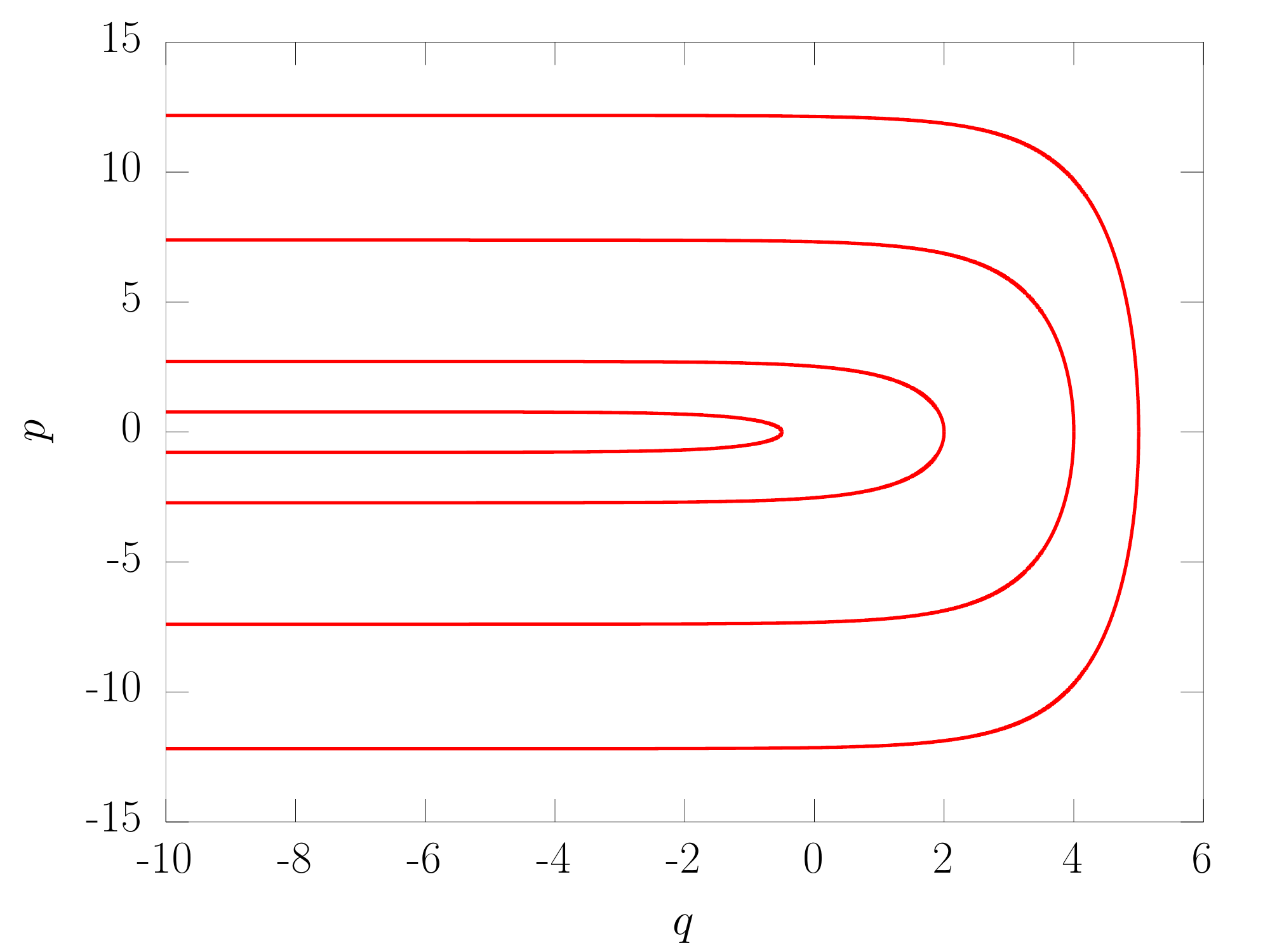}
\par\end{centering}
\caption{Phase portrait of Hamiltonian (\ref{eq:Efefctive_Ham_alpha_zero}).\label{fig:Phase-portrait-of}}

\end{figure}

\noindent We may also check a bit the newton equation (\ref{eq:Newton1})
in this case, we end up with a conservative system with a force deriving
from a potential in one dimension, so we can solve it. We can define
a Hamiltonian 
\begin{equation}
H=\frac{p^{2}}{2}+e^{q}\;,\label{eq:Efefctive_Ham_alpha_zero}
\end{equation}
and find the constant energy trajectories. Given the shape of the
potential energy, we see that for a given total energy $E$ (that
is always positive), we will have a maximum of $q$ attained when
the momentum is $p=0$ at $q^{*}=\log E$, and when $q\rightarrow-\infty$,
we will end up with momenta $p_{\infty}=\pm\sqrt{E}$, so we can sketch
a phase portrait as depicted in Fig.~\ref{fig:Phase-portrait-of},
and can find again the solutions analytically using the first order
ordinary differential equation given by the Hamiltonian (\ref{eq:Efefctive_Ham_alpha_zero}).

\subsection{Solution for $a\protect\ne0$}

We had left with 
\begin{equation}
\ddot{\varphi}=-e^{\varphi}-e^{2t}\:,\label{eq:Newton1-1-1-1}
\end{equation}
We can directly obtain a Hamiltonian 
\begin{equation}
H=\frac{p^{2}}{2}+e^{q}\pm qe^{2t}\:.\label{eq:Hamiltno_autocoherent-1}
\end{equation}
but we may also $\delta\varphi=\varphi-2t$, we end up with 
\begin{align*}
\ddot{\delta\varphi} & =-e^{\delta\varphi+2t}-e^{2t}\:\\
 & =-e^{2t}(1+e^{\delta\varphi})
\end{align*}
So we can summarize the dynamics of the system with a different Hamiltonian
of the type
\begin{equation}
H=\frac{p^{2}}{2}+e^{2t}(e^{q}\pm q)\:.\label{eq:Hamiltno_autocoherent}
\end{equation}
Unfortunately, this is a time-dependent Hamiltonian, and as such not
easy to resolve. Similar Hamiltonian with an exponential growth of
the amplitude of the potential appears in the literature like for
instance in systems dealing with wave particle interactions \citet{Benisti07,Benisti2015}, however
the techniques applied there do not seem to be directly applicable,
so in order to look at the behavior of the potential we need to resort
to a numerical resolution.

\subsection{Numerical solutions for $a\protect\ne0$}

Looking at Eq.~(\ref{eq:Adimensionalized_self_conis}), we may define
an adimensionalized current $j$ as $j=\exp(\phi)$, since we expect
$j$ to be smooth at the origin $\frac{dj}{dr}(0)=0$, this implies
as well that $\frac{d\phi}{dr}(0)=0$. We shall thus impose this in
our initial conditions.

\begin{figure}
\begin{centering}
\includegraphics[width=8cm]{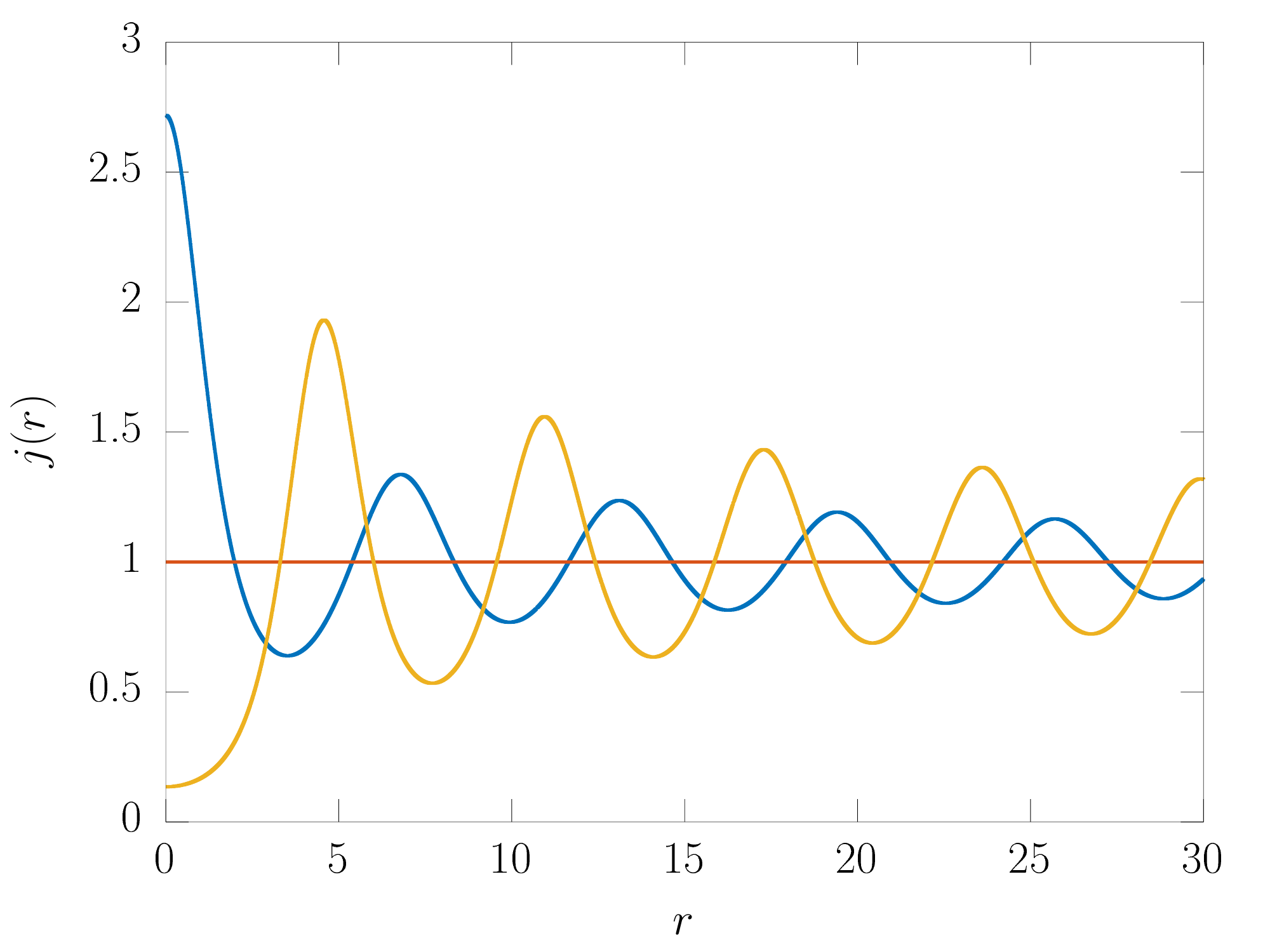}
\par\end{centering}
\caption{Solutions of Eq.(\ref{eq:Adimensionalized_self_conis}) with the minus
sign. The fixed point $j=1$ is an attractor and some current oscillations
are observed. Initial conditions are $\phi(0)=1$ (blue), $\phi(0)=0$ (orange) and
$\phi(0)=-2$ (yellow).\label{fig:Current with minus}}
\end{figure}

We may anticipate that solutions with $a<0$ ($-$ sign in Eq.~(\ref{eq:Adimensionalized_self_conis})
or $a=-1/4$ in Eq.~(\ref{eq:implicit_sans_dim})) are not what we
would expect from a physical behavior. We can as well notice that
$\phi=0$ is a solution (fixed point) that implies as well $j=1$,
so a uniform current. Looking at the numerical solution, we can see
in Fig.\ref{fig:Current with minus}~ 
that the fixed point is actually an attractor and that we observe
current oscillations. The full solution is indeed not realistic at
least when $r\rightarrow\infty$, but it could have some meaning for
small values of $r$.

\begin{figure}
\begin{centering}
\includegraphics[width=8cm]{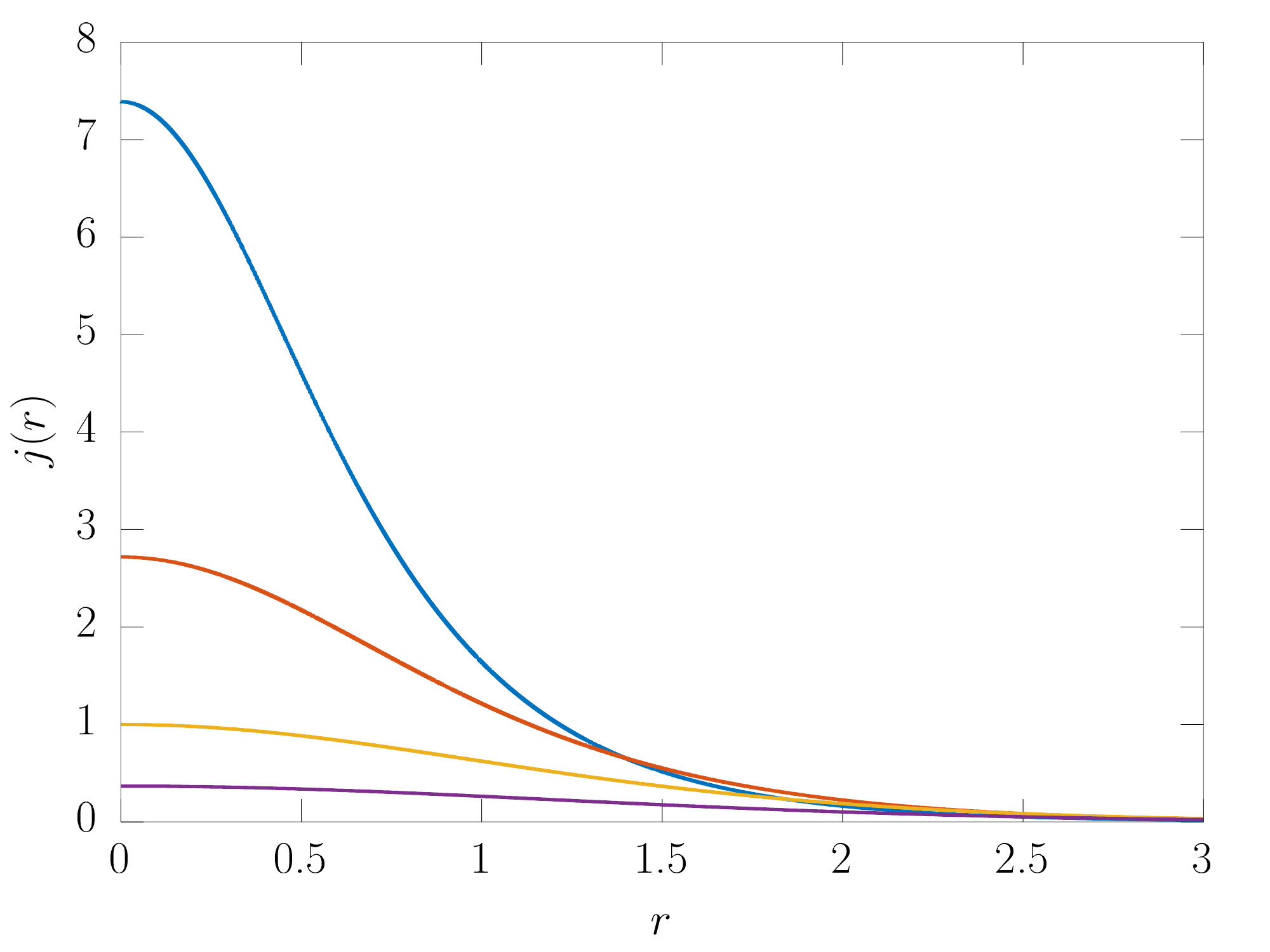}
\par\end{centering}
\caption{Solutions of Eq.(\ref{eq:Adimensionalized_self_conis}) with the plus
sign. All profile are monotonous and decrease quite fast and the fixed
point $j=0$ is an attractor. Initial conditions are $\phi(0)=2$ (blue), $\phi(0)=1$(orange),
$\phi(0)=0$ (yellow) and $\phi(0)=-1$ (magenta).\label{fig:Current with plus}}
\end{figure}

Let us now turn to more relevant physical solutions with $a>0$ ($+$ sign in Eq.~(\ref{eq:Adimensionalized_self_conis}) or $a=1/4$ in Eq.~(\ref{eq:implicit_sans_dim}). 
Solutions are displayed in Fig.~\ref{fig:Current with plus}.
We notice that they are quite regular and no special behavior is observed when the initial condition is changed, we can expect some flattening when $\phi(0)$ becomes small. 
We can from this derive some behavior of $\phi$ assuming $\phi(0)\ll0$, we can then neglect the exponential term in Eq.(\ref{eq:Adimensionalized_self_conis})),
which leads to $\phi(r)\approx-r^{2}/4+\phi(0)$, and we see this implies that $\psi(r)$ becomes some constant, and we can thus explain
the flattening of the current profile.

\begin{figure}
\begin{centering}
\includegraphics[width=8cm]{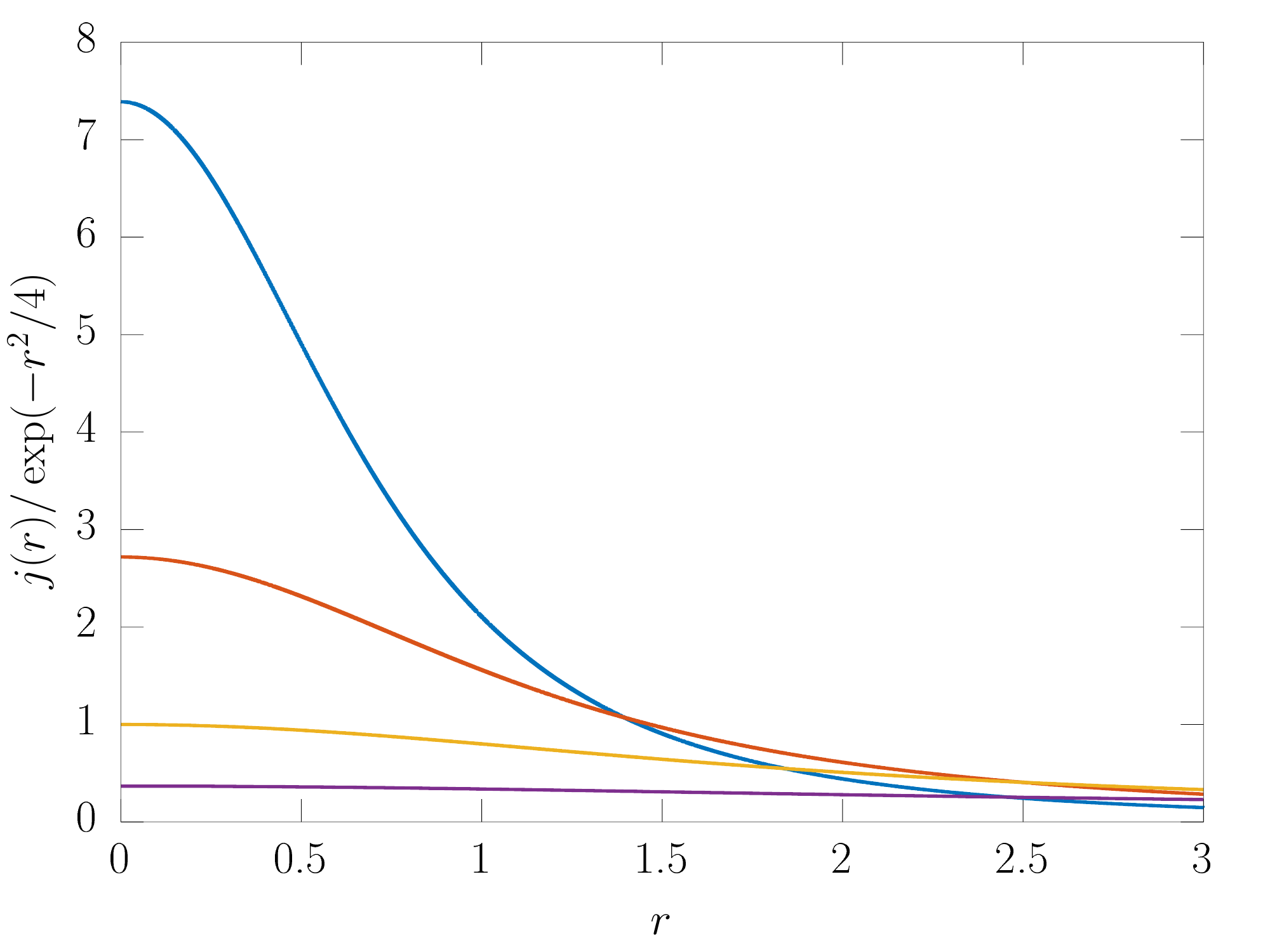}
\par\end{centering}
\caption{Influence of the presence of the non-zero $\gamma_{z}$ in the profile.
We can see that as soon as $\phi(0)>0$, we end up with a better radial
confinement of the plasma. We consider the solutions of Eq.(\ref{eq:Adimensionalized_self_conis})
with the plus sign. Initial conditions are $\phi(0)=2$ (blue), $\phi(0)=1$(orange),
$\phi(0)=0$ (yellow) and $\phi(0)=-1$ (magenta).\label{fig:Current with plus-1}}
\end{figure}

Since the current profile is directly proportional to the density profile of the plasma, we can see that as soon as $\phi(0)\ge0$ that
the presence of a plasma flow along the $z-$axis enhances plasma confinement, this is illustrated in Fig.~\ref{fig:Current with plus-1}.
.
\begin{figure}
\begin{centering}
\includegraphics[width=5cm]{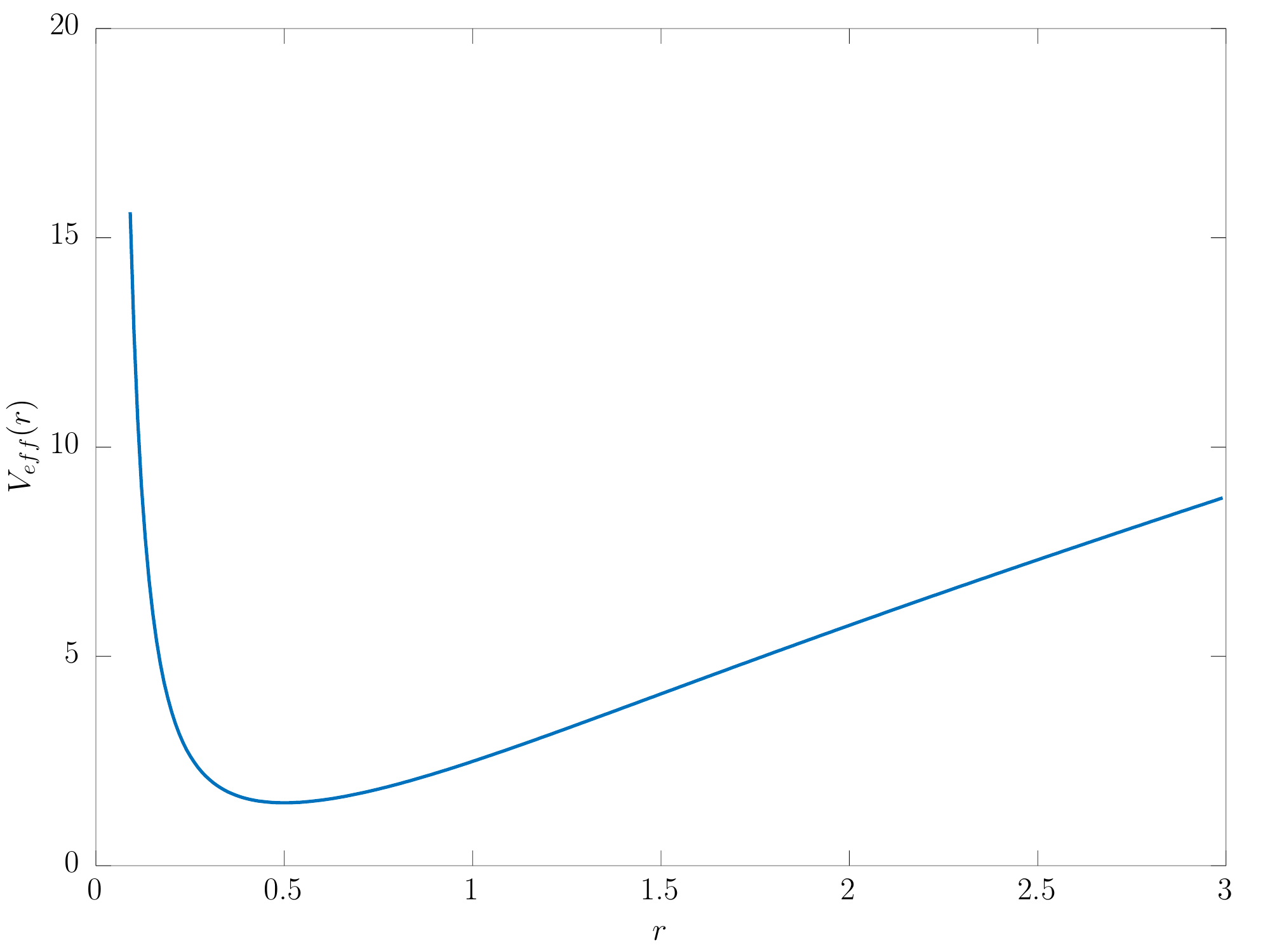}
\includegraphics[width=5cm]{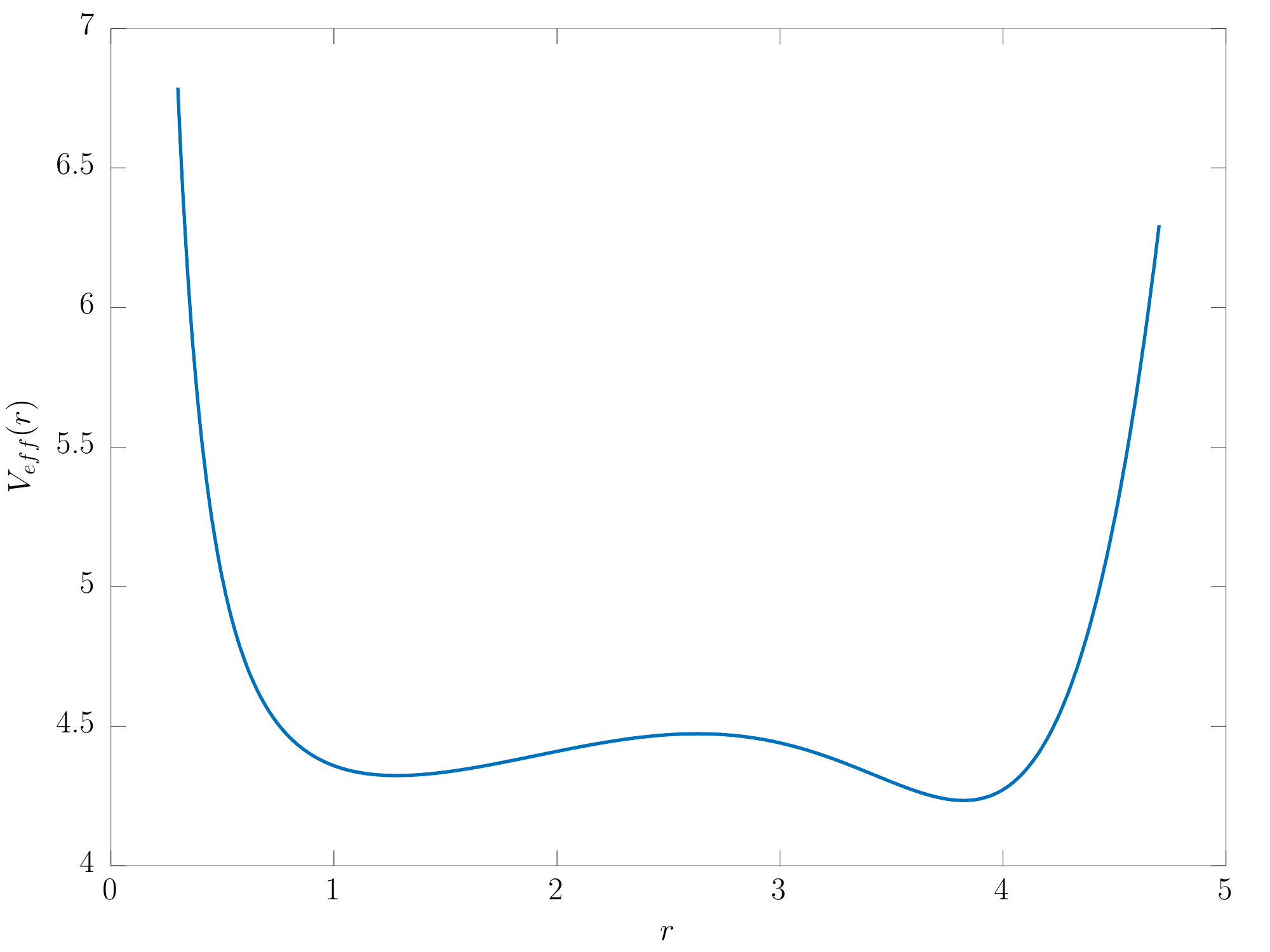}\\
\includegraphics[width=5cm]{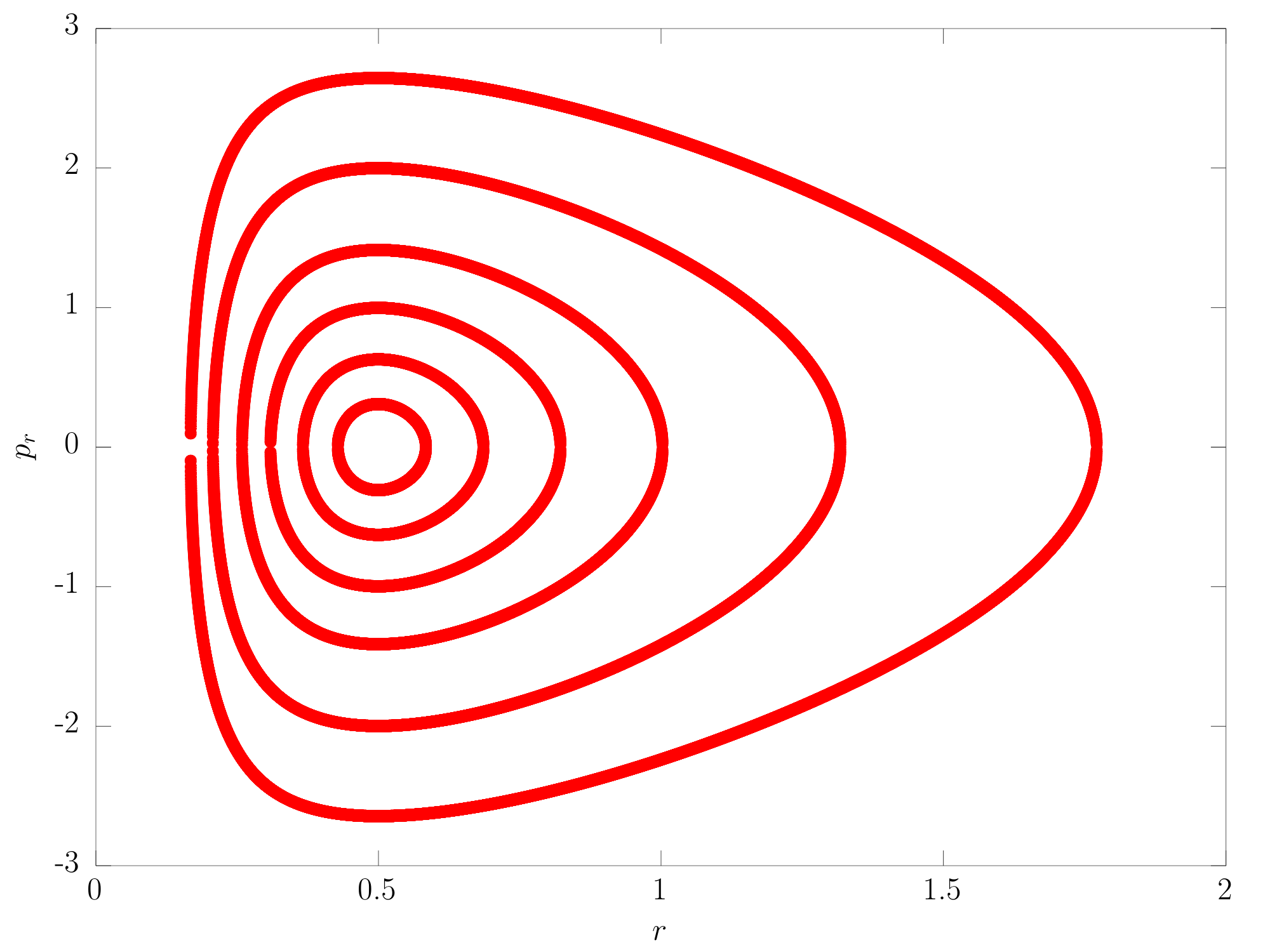}
\includegraphics[width=5cm]{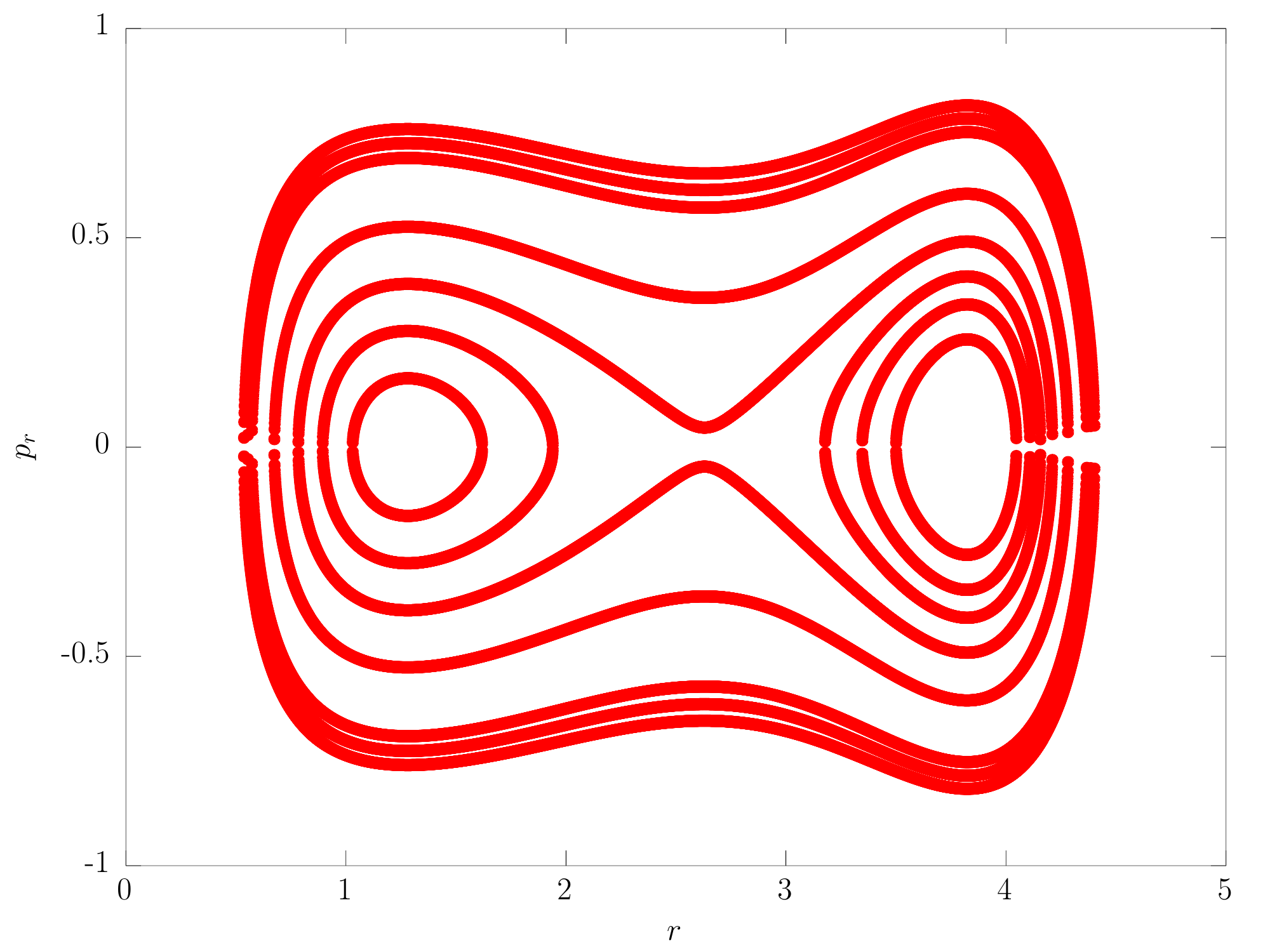}
\par\end{centering}
\caption{Illustration of the effective potential (top) and resulting phase space trajectories (botom) arising with a self-consistent magnetic field. Left: A typical situation when $a>0$ (here $a=1/4$), and  $\phi(0)=2$; the effective potential is computed with $p_\theta=0.5$ and $p_z=3$. Right: A situation when $a<0$ (here $a=-1/4$) and  $\phi(0)=-2$, and unstable arises; the effective potential is computed with $p_\theta=0.5$ and $p_z=-4$. \label{fig:Phase_portrait}}
\end{figure}

When looking for unstable points and separatrices in in the phase
portrait of the integrable Hamiltonian (\ref{eq:Hamiltonien}), with
a self-consistent component of the magnetic field, we found out that
the solutions with $a>0$ are not able to create a strong enough modulation
to trigger the presence of an unstable point. Given the conclusions
found in \citet{Cambon2014}, we thus expect that these configurations should lead as
well to regular trajectories when going to a toroidal configuration,
at least in the large aspect ratio limit and this hints towards some
stability of these solutions. On the contrary when $a\le0$ it is
possible to find self-consistent configurations that lead to the presence
of a separatrix, and thus radial chaotic motion of particles can be
expected if such configurations locally appear in the plasma, the phenomena
are illustrated in Fig.~\ref{fig:Phase_portrait}.

\section{Conclusion}

In this paper we have built self-consistent solutions of the Vlasov equation in a cylindrical magnetized plasma, these solutions correspond
to a thermodynamic equilibrium of a non self-consistent ``passive'' plasma, and as such can potentially accommodate with a different thermodynamic
temperatures of the ions and the electrons. 
Besides the possibility to use these exact solutions as test beds for large numerical kinetic simulations and code validation like for instance GYSELA \citet{Grandgirard2006}, we found out that these solutions
confirm the results proposed in \citet{Ogawa2019}, and that a better radial confinement of the plasma can be obtained by the existence of a longitudinal flow 
(i.e a toroidal flow in a toroidal machine) above a certain threshold.
It could be therefore interesting to check if there is an experimental correlation of these findings, namely if a better confinement of the
plasma in magnetized fusion devices can not be obtained by generating a toroidal momentum of the plasma while heating it with neutral injections beams. 
Finally, our findings show that equilibrium configurations leading to a globally confined plasma ($a>0$) are unable to create unstable points and associated separatrices in the radial direction,
which hints at some stability of these magnetic profiles when considering a confinement within a torus and no destruction of the adiabatic invariant,
at least in the large aspect ratio limit, and as long as electric field effects can be neglected, and this is thus compatible with results discussed in \citet{Sharma83} regarding special Bennett type solutions. 
As a perspective of this work, we may now consider to take into account global diamagnetic effects and see the possible influence of poloidal flow and current on the confinement. This appears as a crucial point,
as when we go back to the toroidal geometry, the destruction of one symmetry will inevitably imply a coupling between both toroidal and poloidal flows of the plasma. 
So the influence of a poloidal flow on confinement needs as well to be addressed in a self-consistent manner and some comparison with results obtained in the MHD framework
\citet{White2011,White2012} could be as well envisioned.

\acknowledgments{This work has been carried out within the framework of the French Research Federation for Magnetic Fusion Studies. The project leading to this publication (TOP project) has received funding from Excellence Initiative of Aix-Marseille University - A*MIDEX, a French ``Investissements d'Avenir'' programme.}
\appendix
\section{Density and current computation}
\unskip
Once the particle density function is obtained, in order to compute quantities such as the particle density or the current
we rewrite $-\beta H-\gamma_{z}p_{z}-\gamma_{\theta}p_{\theta}-\gamma_{0}$
(with $q=m=1$) as

\begin{multline}
-\frac{\beta}{2}\left[p_{r}^{2}+\left(\frac{p_{\theta}}{r}-\left(\frac{B_{0}}{2}-\frac{\gamma_{\theta}}{\beta}\right)r\right)^{2}+\left(p_{z}-\left(A_z(r)-\frac{\gamma_{z}}{\beta}\right)\right)^{2}\right]\\
- \frac{\gamma_{\theta}}{2}(B_{0}-\frac{\gamma_{\theta}}{\beta})r^{2}-\gamma_{z}A_z(r)+\frac{\gamma_{z}^{2}}{2\beta}-\gamma_{0}\:,
\end{multline}
and then just make some Gaussian integrals.


\begin{thebibliography}{10}

\bibitem{Wolf2003}
R.~C. Wolf.
\newblock {Internal transport barriers in tokamak plasmas}.
\newblock {\em Plasma Phys. Control. Fusion}, 45:R1, 2003.

\bibitem{Connor2004}
J.~W. Connor, T.~Fukuda, X.~Garbet, C.~Gormezano, V.~Mukhovatov, M.~Wakatani,
  and {and ITPA Topical Group on Transport and Internal Barrier Physics} {ITB
  Database Group}.
\newblock {A review of internal transport barrier physics for steady-state
  operation of tokamaks}.
\newblock {\em Nucl. Fusion}, 44:R1, 2004.

\bibitem{Balescu98}
R.~Balescu.
\newblock {Hamiltonian nontwist map for magnetic field lines with locally
  reversed shear in toroidal geometry}.
\newblock {\em Phys. Rev. E}, 58:3781, 1998.

\bibitem{Firpo98}
M.-C. Firpo.
\newblock {Analytic estimation of the Lyapunov exponent in a mean-field model
  undergoing a phase transition}.
\newblock {\em Phys. Rev. E}, 57:6599, 1998.

\bibitem{Ogawa2016_2}
S.~Ogawa, X.~Leoncini, G.~Dif-Pradalier, and X.~Garbet.
\newblock {Study on creation and destruction of transport barriers via
  effective safety factors for energetic particles}.
\newblock {\em Phys. Plasmas}, 23:122510, 2016.

\bibitem{Ogawa2019}
Shun Ogawa, Xavier Leoncini, Alexei Vasiliev, and Xavier Garbet.
\newblock {Tailoring steep density profile with unstable points}.
\newblock {\em Phys. Lett. A}, 383:35--39, 2019.

\bibitem{Cambon2014}
B.~Cambon, X.~Leoncini, M.~Vittot, R.~Dumont, and X.~Garbet.
\newblock {Chaotic motion of charged particles in toroidal magnetic
  configurations}.
\newblock {\em Chaos}, 24:033101, 2014.

\bibitem{Neishtadt86}
A.~I. Neishtadt.
\newblock {On the change of adiabatic invariant due to separatrix crossing}.
\newblock {\em Sov. Phys. Plasma Phys.}, 12:568--573, 1986.

\bibitem{Tennyson86}
J.~Tennyson, J.~R. Cary, and D.~F. Escande.
\newblock {Change of the Adiabatic Invariant due to Separatrix Crossing}.
\newblock {\em Phys. Rev. Lett.}, 56:2117--2120, 1986.

\bibitem{Ogawa2016}
S.~Ogawa, B.~Cambon, X.~Leoncini, M.~Vittot, D.~{Del Castillo-Negrete},
  G.~Dif-Pradalier, and X.~Garbet.
\newblock {Full particle orbit effects in regular and stochastic magnetic
  fields}.
\newblock {\em Phys. Plasmas}, 23:072506, 2016.

\bibitem{Leoncini2018}
Xavier Leoncini, Alexei Vasiliev, and Anton Artemyev.
\newblock {Resonance controlled transport in phase space}.
\newblock {\em Physica D: Nonlinear Phenomena}, 364:22--26, 2018.

\bibitem{Jeans1915}
J.~H. Jeans.
\newblock {On the theory of star-streaming and the structure of the universe}.
\newblock {\em Mon. Not. Roy. Astro. Soc.}, 76:70--84, 1915.

\bibitem{Vlasov38}
A.~A. Vlasov.
\newblock {The vibrational properties of an electron gas}.
\newblock {\em Zh. Eksp. Ther. Fiz.}, 8:291, 1938.

\bibitem{Vlasov68}
A.~A. Vlasov.
\newblock {\em Sov. Phys. Uspekhi}, 93, 1968.

\bibitem{Binney2008}
J.~Binney, S.~Tremaine, et~al.
\newblock {\em {Galactic dynamics}}.
\newblock Princeton University Press, 2008.

\bibitem{Jaynes57}
E.~T. Jaynes.
\newblock {Information Theory and Statistical Mechanics}.
\newblock {\em Phys. Rev.}, 106:620, 1957.

\bibitem{Zubarev1996}
D.~Zubarev, V.~Morozov, and G.~R{\"o}pke.
\newblock {\em {Statistical Mechanics of Nonequilibrium Processes}}.
\newblock Akademie Verlag GmbH, 1996.

\bibitem{Yamaguchi04}
Y.~Y. Yamaguchi, J.~Barr{\'e}, F.~Bouchet, T.~Dauxois, and S.~Ruffo.
\newblock {Stability criteria of the Vlasov equation and quasi-stationary
  states of the HMF model}.
\newblock {\em Physica A}, 337, 2004.

\bibitem{Bennett34}
W.~H. Bennett.
\newblock {Magnetically Self-Focussing Streams}.
\newblock {\em Phys. Rev.}, 45:890, 1934.

\bibitem{Bennett55}
W.~H. Bennett.
\newblock {Self-Focusing Streams}.
\newblock {\em Phys. Rev.}, 98:1584, 1955.

\bibitem{Cecherini05}
Ceccherini, F.; et al
\newblock {Two-dimensional Harris-Liouville plasma kinetic equilibria}
\newblock {\em Phys. Plasmas} {\bf 2005}, {\em 12},~052506

\bibitem{Benisti07}
D.~Benisti and L.~Gremillet.
\newblock {Nonlinear plasma response to a slowly varying electrostatic wave,
  and application to Stimulated Raman Scattering}.
\newblock {\em Phys. Plasmas}, 14:042304, 2007.

\bibitem{Benisti2015}
Didier B{\'e}nisti and Laurent Gremillet.
\newblock {Global change in action due to trapping: How to derive it whatever
  the rate of variation of the dynamics}.
\newblock {\em Phys. Rev. E}, 91:042915, 2015.

\bibitem{Grandgirard2006}
V.~Grandgirard, M.~Brunetti, P.~Bertrand, N.~Besse, X.~Garbet, P.~Ghendrih,
  G.~Manfredi, Y.~Sarazin, O.~Sauter, E.~Sonnendr{\"u}cker, J.~Vaclavik, and
  L.~Villard.
\newblock {A drift-kinetic Semi-Lagrangian 4D code for ion turbulence
  simulation}.
\newblock {\em Journal of Computational Physics}, 217(2):395--423, 2006.

\bibitem{Sharma83}
A.~S. Sharma.
\newblock {Vlasov Stability of Bennett Equilibrium}.
\newblock {\em Nucl. Fusion}, 23:1493, 1983.

\bibitem{White2011}
R.~B. White.
\newblock {Modification of particle distributions by MHD instabilities II}.
\newblock {\em Plasma Phys. Control. Fusion}, 53:085018, 2011.

\bibitem{White2012}
R.~B. White.
\newblock {Modification of particle distributions by MHD instabilities I}.
\newblock {\em Commun. Nonlinear Sci. Numer. Simulat.}, 17:2200, 2012.

\end{thebibliography}
\end{document}